\documentclass[preprint,5p]{elsarticle}
\usepackage{graphicx}
\usepackage{amssymb,amsmath,mathrsfs}
\journal{Physics Letter B}

\begin{document}
\begin{frontmatter}
\title{On the impact of large amplitude pairing fluctuations on nuclear spectra }

\author[UAM]{Nuria L\'opez Vaquero}
\author[GS]{Tom\'as R. Rodriguez}
\author[UAM]{J. Luis Egido}
\ead{j.luis.egido@uam.es}

\address[UAM]{Departamento de F\'isica Te\'orica, Universidad Aut\'onoma de Madrid, E-28049 Madrid, Spain}
\address[GS]{Gesellschaft f\"ur Schwerionenforschung (GSI), D-64291 Darmstadt, Germany}

\begin{abstract}
  The influence of large amplitude pairing fluctuations is investigated in the framework of beyond mean
field symmetry  conserving configuration mixing calculations.  In the numerical application the finite range
density dependent Gogny force is used. We investigate the nucleus $^{54}$Cr with particle number and angular
momentum projected wave functions considering the axial quadrupole deformation and the pairing gap degree of freedom
as generator coordinates.  We find that the  effects of the pairing fluctuations  increase with the excitation energy
and the angular momentum. The self-consistency in the determination of the basis states plays an important role.

\end{abstract}


\end{frontmatter}

It is well known that pairing correlations play a predominant role in most nuclear phenomena \cite{BMP58,RS80}. In spite of that nuclear superfluidity is not completely understood  and it is still 
a very active research field \cite{LE09,POT10,Pi10,GE11}.  
Nowadays pairing correlations are routinely included in all mean field nuclear structure calculations either in the Hartree-Fock (HF) plus BCS
approach or in the Hartree-Fock-Bogoliubov (HFB) approach. However while the HFB (BCS) approach is very successful to describe infinite systems, it was soon recognized \cite{RS80}  that it is a deficient theory in the case of  atomic nuclei where only a few pairs of nucleons (weak pairing regime) participate in the superfluid phase. Furthermore in a finite system  a sharp number of particles is needed.

In spite of this fact the modern beyond mean field theories (BMFT) that have been developed in recent years \cite{Bender_RMP_03}  do not treat pairing on the same footing as the nuclear deformation. From those theories the so called Symmetry Conserving Configuration Mixing (SCCM) approaches have played a major role. Their two basic ingredients are:  first  the restoration of the symmetries broken at the mean field level, i.e., the angular momentum projection -and sometimes particle  number projection- and  the second one  the configuration mixing. 
The latter one is done in the Generator Coordinate Method (GCM)  which allows to deal easily with collective degrees
of freedom. The difficulty to solve the corresponding variational equations, however,  increases considerably with the number of coordinates and until now only the deformation parameter $\beta$  has been considered in axially symmetric calculations \cite{Bender_Kr_06,RODGUZNPA02,Niksic_PNAMP_06,Rod_CaTiCr_07} and $(\beta,\gamma )$  and $K$ mixing in triaxial ones \cite{BendHFB_SK_08,RingGCM_Rel_10,RE_10}.   It is the purpose of this letter  to consider explicitly for the first time the collective pairing degree of freedom with a finite range force in the framework of the SCCM\footnote{A list with the  introduced acronyms is given at the end of the Letter}. approach with axially symmetric shapes.  

Some arguments indicating the relevance of  considering the pairing fluctuations are:
First,  the monopole (pairing) and the quadrupole (deformation)  are the most relevant degrees of freedom, it seems therefore reasonable to consider both of them on an equal footing.  Second, the constrained variational principle of Ritz used to determine the intrinsic basis states is very effective in determining the wave function (w.f.) of the ground state with the given quantum numbers  and constraints.   Ground states of the SCCM calculations benefit of this fact, however, SCCM excited states with the same or different quantum numbers are not favored by it and depend more strongly on the basis size (number of generator coordinates). Consequently, in restricted self-consistent calculations  a stretched spectrum is forecasted which will be  squeezed by an appropriate increase of the basis size, for example by allowing pairing fluctuations.  Lastly, the pairing vibrations on their own are very interesting and a simultaneous study on shape and pairing fluctuations will allow us to disentangle if there exist genuine pairing vibrations or alternatively to which degree they are damped.        

    In the gauge space associated with pairing the HFB w.f.  has two collective degrees of freedom, the pairing
gap $\Delta$, which measures the amount of  pairing correlations, i.e., the ``deformation'' \cite{Brog_rev} in the associated gauge space, and  the angle $\varphi$ which indicates  the orientation of the HFB state in this space. The HFB minimization determines  the w.f. and thereby $\Delta$   while the gauge angle $\varphi$ does not play any role at the mean field level.  The degree of freedom associated  to $\varphi$  has been exploited in the past  \cite{DMP64}:  linear combinations of w.f.'s with different orientation in the gauge space  provide a number conserving wave function.  Pairing vibrations -associated with w.f.'s with different pairing gaps- around the average gap parameter $\Delta_0$ of the energy minimum, on the other hand, have attracted little attention. As a matter of fact they have been considered only either with very schematic interactions in the framework of the collective Hamiltonian \cite{Bes_coll,Pomo_85}, in  microscopic model calculations \cite{Mafer03,Mafer05}, in reduced configuration space \cite{Faess_73} or in earlier BMFT approaches \cite{Meyer,Heenen}.

 To quantify the pairing content of a w.f. with a finite range interaction,  like the Gogny force, that provides
state dependent gaps is not trivial.  A quantity that supplies a measure of the pairing correlations and is easy to handle  is the  mean square deviation of the particle number  $(\Delta\hat{N})^2$. This quantity is zero in the absence of pairing correlations and is large for strongly correlated systems. Furthermore, since for a schematic pairing interaction $\langle(\Delta \hat{N})^2\rangle= 4 \sum_{k>0} u_k^2 v_k^2= \Delta^2\sum_{k>0} \frac{1}{E_k^2}$, with $E_k$
the quasiparticle energy,   $ \Delta \propto \langle(\Delta \hat{N})^2\rangle^{1/2}$ and  $(\Delta \hat{N})^2$ provides an indication of the pairing content of the wave function. In the following we will
denote  $\delta = \langle\phi|(\Delta \hat{N})^2|\phi\rangle^{1/2}$ and use it as coordinate to generate wave functions with different pairing
correlations. Of course one could constrain separately the mean square deviation for protons and for neutrons, this is more complicated  and work in this direction is on progress. In this first exploratory letter we constrain only  the total mean square deviation.
 
In order to implement pairing fluctuations together with axially symmetric quadrupole fluctuations we proceed in the following way:  First, we generate   
intrinsic HFB wave functions $|\phi(q,\delta)\rangle$ with given quadrupole deformation $q$ and  ``pairing deformation'' $\delta$ by solving the variational equation
 \begin{equation}
 \delta {E^{\prime}}[\phi(q,\delta)]   = 0  \label{min_E},
\end{equation}
with 
\begin{equation}
{E^{\prime}}= \frac{\langle\Phi|\hat{H}|\Phi \rangle} {\langle\Phi|\Phi \rangle}   
 - \lambda_q \langle \phi |\hat{Q}_{20} | \phi \rangle- \lambda_\delta \langle \phi|(\Delta\hat{N})^2|\phi \rangle^{1/2}, \label{E_prime}
  \end{equation}
 and the Lagrange multipliers $\lambda_q$ and $\lambda_\delta$ being determined by the constraints  
 \begin{equation}
 \langle \phi |\hat{Q}_{20} | \phi \rangle =q, \;\;\;\; \;\;\;\;      \langle \phi|(\Delta\hat{N})^2|\phi \rangle^{1/2} = \delta,
 \end{equation}
 $\hat{Q}_{20}$  is the quadrupole mass operator.
If in Eq.~\ref{E_prime}, $|\Phi\rangle\equiv |\phi \rangle$  we are solving the plain HFB equations\footnote{ In this case
we have to add to Eq.~\ref{E_prime}  a term $-\lambda_{N} \hat{N}$ to keep the particle number right on the average, with $\lambda_{N}$ fixed by the constraint $\langle\phi | \hat{N}| \phi\rangle=N $.}.  A Particle Number Projection (PNP)  and/or  Angular Momentum Projection (AMP)  out of this w.f. would be a Projection After Variation (PAV). 
However if $|\Phi\rangle\equiv \hat{P}^N|\phi \rangle$, being $\hat{P}^N$ the particle number projector, the determination of 
$|\phi\rangle$ is done in the so called Variation After (Particle Number) Projection (PN-VAP) approach. This method provides a much better description of the pairing correlations in the intrinsic w.f.  \cite{Anguiano_VAP_02} although is more involved. There are only few implementations of the PN-VAP approach using either separable forces \cite{EGPNP}, small configuration spaces  \cite{Carlo} or the most recent ones with the Gogny  \cite{AER.01P} and Skyrme functionals \cite{STOITSOV, LACROIX}. Finally, as in the HFB case, an angular momentum projection can be performed afterwards. The variational equations are solved using the conjugate gradient method \cite{grad}.
   Once we have generated the basis states we can proceed with the configuration mixing calculation. This
    is carried out within the generator coordinate method  taking linear combinations of the wave functions obtained in the first step after performing projections on the required symmetries
\begin{eqnarray}
|\Psi^{I,\sigma}\rangle=\int f^{I,\sigma}(q,\delta) \; \hat{P}\;|\phi(q,\delta)\rangle dq d\delta.
\label{GCMstate}
\end{eqnarray} 
The variational principle applied to the weights $ f^{I,\sigma}(q,\delta)$ gives the generalized eigenvalue problem,  the Hill-Wheeler (HW) equation:
\begin{eqnarray}
\!\!\!\!\!\!\!\!\!\!\!\!\!\!\!\int \left(\mathcal{H}^{I}(q \delta,q^\prime \delta^\prime)-E^{I,\sigma}\mathcal{N}^{I}( q \delta,q^\prime \delta^\prime)\right)f^{I,\sigma}(q^\prime \delta^\prime)dq^\prime d\delta^\prime=0,
\label{HWeq}
\end{eqnarray}
with  $\mathcal{H}^{I}$ and $\mathcal{N}^{I}$ the  Hamiltonian and norm  overlaps, respectively, see \cite{RODGUZNPA02} for further details.  The symbol $\sigma=1,2, ...$, numbers the ground and excited states with angular momentum $I$, $E^{I,\sigma}$ is the energy of the corresponding state.
If in Eq.~\ref{GCMstate} the projector $\hat{P}= \hat{P}^Z\hat{P}^N\hat{P}^I$,  with $ \hat{P}^{I}$  a short-hand notation for the angular momentum projection operator,
we are performing a PNP and AMP.    In previous calculations  \cite{RODGUZNPA02}  \cite{Niksic_PNAMP_06} one can find configuration mixing calculations with only AMP, in this case $\hat{P}= \hat{P}^I$. We shall also discuss below such kind of calculations.

\begin{figure}[t]
{\centering {\includegraphics[angle=0,width=0.99\columnwidth]{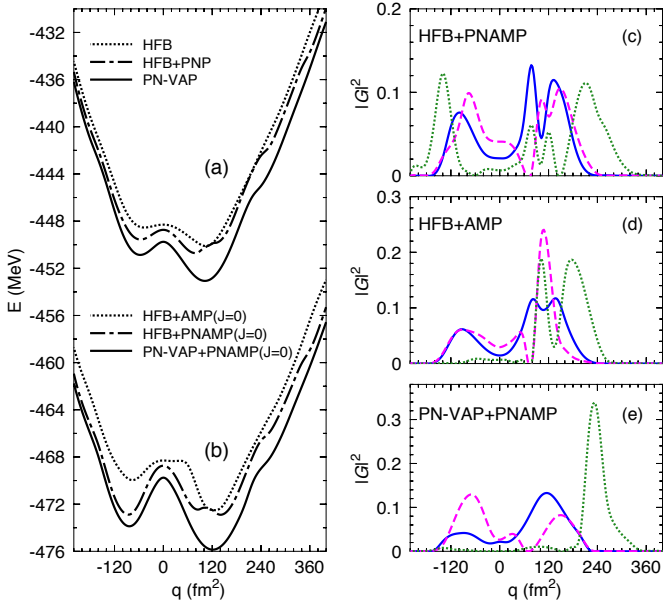}} \par}
\caption{(Color online) Left: Potential energy curves as a function of the quadrupole moment for the nucleus $^{54}$Cr in various approaches.
The energies of (a) have been shifted by 20 MeV. Right: The wave functions of the lowest $0^+$ collective states 
($0^+_1$ continuous line, $0^+_2$ dashed line and $0^+_3$ dotted line ) in the  corresponding approaches.   }\label{q_plots}
\end{figure}

We use in our numerical application the finite range density dependent Gogny force
with the D1S parametrization \cite{BERGNPA84}, which is well known for its successful predictions. In the configuration space we take eight oscillator shells;
for the $q$ coordinate we take an interval $-220$ fm$^2$ up to $+400$  fm$^2$ with  $\Delta q=20$ fm$^2$ and  for the pairing coordinate $\delta$ from $0$ 
to $4.5$ with a $0.5$ step size. The chosen $\delta$ interval covers a pairing energy range from  $0.0$ up to $~\sim 50$ MeV, to compare with typical values of a few MeV's in  the  nucleus we shall analyze.
In both cases the intervals have been chosen in such a way that an  energy convergence is obtained and that the tails of the collective wave functions go to zero inside the interval border.

  \begin{figure}[t]
{\centering {\includegraphics[angle=0,width=0.99\columnwidth]{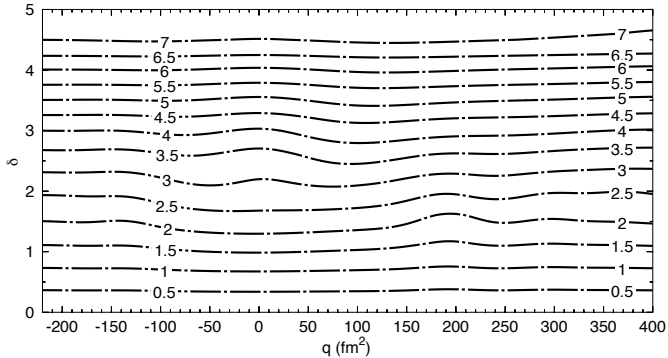}} \par}
\caption{Contour plots of the square root of the absolute value of the pairing energy in MeV in the $(q,\delta)$ plane.}\label{cont_pair_ener}
\end{figure}

 As an example we will apply our theory to the nucleus  $^{54}$Cr. Our first intention is to investigate  a numerical
 case to quantify the relevance of the pairing fluctuations and not to make an exhaustive comparison with the experiment. We are aware that relevant degrees of freedoms like triaxiality,
 or the use of a larger configuration space to describe very excited states,  are necessary for such a goal. A comparison with the experiment for the $2^+$ states in this region without pairing fluctuations can be found in ref.~\cite{Rod_CaTiCr_07}.
   In order to disentangle the effects of particle number projection we will present three different methods, namely,  HFB+AMP, HFB+PNAMP and PN-VAP+PNAMP to determine the w.f.'s $\hat{P} |\phi(q,\delta)\rangle$ of Eq.~\ref{GCMstate}, see Table~\ref{Table1}.     

  \begin{table}[t]
\begin{center}
\begin{tabular}{|c|c|c|}\hline
  Name of the & Intrinsic w.f. $|\phi \rangle$ fixed &  $\hat{P}$ \\ 
   Approach                       &  by minimization of: &\\ \hline
HFB+AMP & ${\langle\phi|\hat{H}|\phi \rangle} /{\langle\phi|\phi \rangle}$ & $\hat{P^{I}}$ \\\hline
HFB+PNAMP & ${\langle\phi|\hat{H} |\phi \rangle}/ {\langle\phi|\phi \rangle}$ & $\hat{P^{I}} \hat{P^{N}}$ \\\hline
PN-VAP+PNAMP & ${\langle\phi|\hat{H}\hat{P^{N}}|\phi \rangle}/ {\langle\phi|\hat{P^{N}} |\phi \rangle}$ & $\hat{P^{I}} \hat{P^{N}} $ \\\hline  
\end{tabular}
\end{center}
\caption{Approaches used to calculate the w.f. $\hat{P}|\phi(q,\delta)\rangle$ of  Eq.~\ref{GCMstate}: In the name of each approach the letters before the plus sign indicate the way in which the intrinsic w.f. $|\phi(q,\delta)\rangle$  is determined. The letters after the plus sign indicate the symmetries we are considering in the projection $\hat{P}$.}
\label{Table1}
\end{table}
In the first one (HFB+AMP) particle number projection is completely ignored, i.e., the intrinsic w.f., $|\phi(q,\delta) \rangle$, is determined in the HFB approach and afterwards AMP is performed. In the second one (HFB+PNAMP) $|\phi(q,\delta) \rangle$ is determined in the HFB approach but then PNP and AMP are performed. In the last one (PN-VAP+PNAMP) $|\phi(q,\delta) \rangle$ is determined in the PN-VAP method and afterwards the PNP and AMP are performed. This w.f.'s can be used by their own (to calculate potential energy surfaces for example) or as basis states of the SCCM calculations, i.e., to solve Eq.~\ref{HWeq}; we will present examples of both cases. See also Table\ref{Table1}. Furthermore in order to investigate the effect of the pairing fluctuations we shall  present results in one dimension (1D), i.e., only with the quadrupole moment as coordinate generator and in two dimensions (2D) with the pairing energy and the quadrupole moment as generator coordinates.

 Prior to the SCCM calculations we  present in Fig.~\ref{q_plots}(a)  an example of potential energy curves as a function of the quadrupole moment (1D) calculated within the three approximations described above for the nucleus $^{54}$Cr. 
All the three calculations present two minima, one prolate and one oblate, being the HFB energy (dotted line) the flattest one. The  HFB+PNP wave functions  (dash-dotted line) produces an additional energy lowering with the exception of some values around  $q\sim120$ fm$^2$ where the pairing correlations are zero. This is not the case in the PN-VAP approach 
where a larger energy gain is obtained for all $q$ values. In Fig.~\ref{q_plots}(b) we display the AMP curves for $I=0 \; \hbar$ where we can observe the effects of the angular momentum projection (additional energy gain and slight displacement of the minima to higher deformations). Now we proceed to the SCCM approaches:  In Fig.~\ref{q_plots}(c)-(e) we show the wave functions of the ground and first two excited states for $I =0 \hbar$ obtained by solving Eq.~\ref{HWeq} for the different cases. These collective wave functions represent the distribution of probability of having the states with a given intrinsic deformation\footnote{See \cite{RS80} for the relationship of the weights $G$ used to plot the w.f. and the weights $f$ of Eq.~\ref{GCMstate}.}.
We discuss the SCCM results for the PN-VAP+PNAMP case as an example. We observe in Fig 1(e) that the structure of the states is strongly correlated to the shape of the potential energy curve. The $0^+_1$ state presents two peaks at $q=120$ fm$^2$ (the largest)  and $q=-80$ fm$^2$, the $0^+_2$ state has also two peaks but this time the oblate one is the largest one. Finally, the $0^+_3$ state presents a narrow peak structure at $q=240$ fm$^2$ corresponding to the shoulder of the potential energy curve at this deformation. 
  
  \begin{figure*}[t]
\begin{center}
\includegraphics[angle=0,scale=.7]{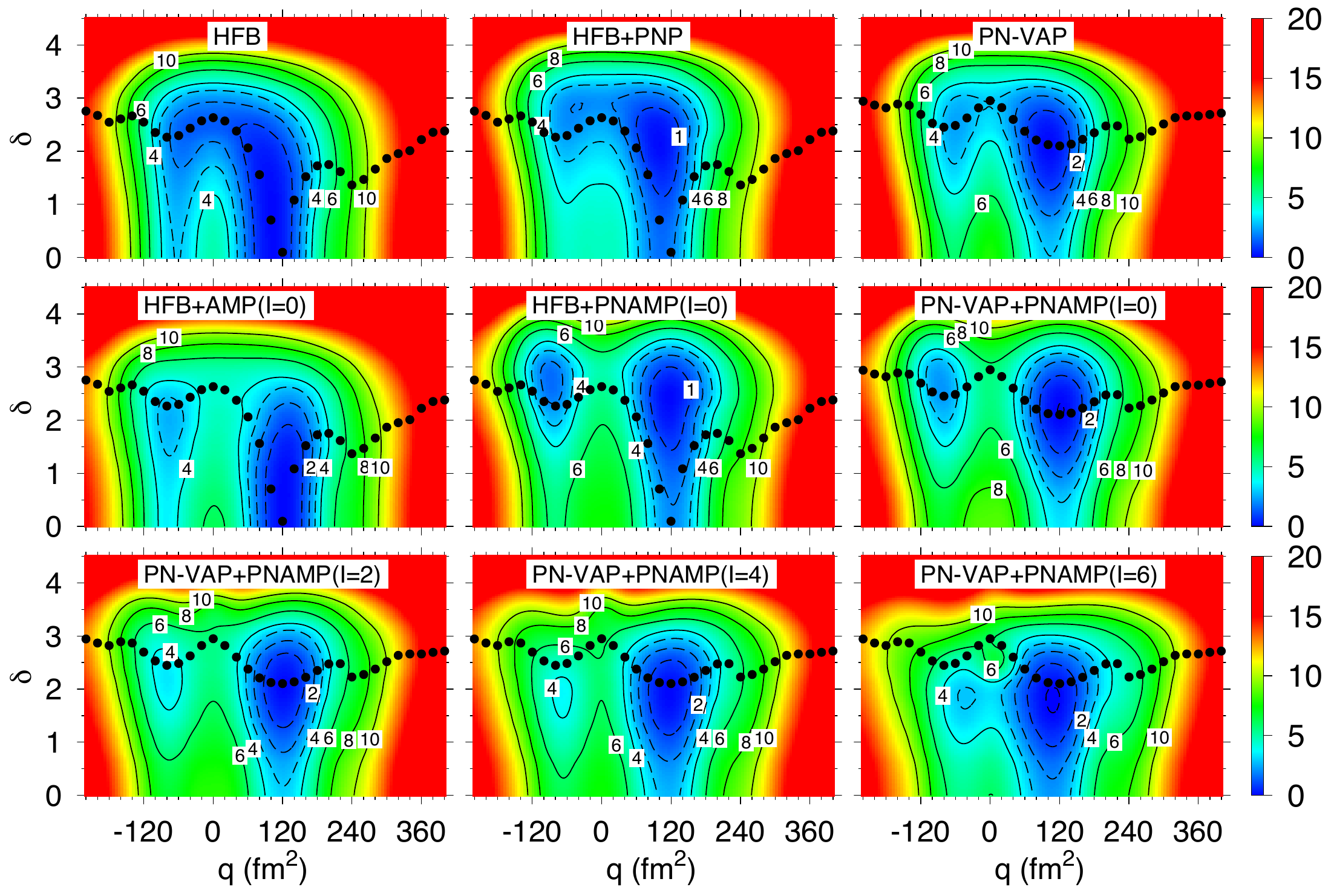}
\end{center}
\caption{(Color online) Energy contour plots in the plane $(\delta,q)$ in different approaches for $^{54}$Cr, see main body for further explanations.  The dashed lines represent contours from
0 to 3 MeV in 1MeV steps. The continuous lines represent contours from 4 to 10 MeV in 2 MeV steps. The energy origin has been chosen independently
for each panel and the energy of the minimum has been set zero. The energy scale on the right is in MeV.
}\label{2_dim_54Cr}
\end{figure*} 
  
 We now discuss the inclusion of the $\delta$ degree of freedom. In Fig.~\ref{cont_pair_ener} we present contour lines of the square root of the pairing energy
 in the $(q, \delta)$ plane for the wave functions of the PN-VAP+PNAMP  approach for $I=0 \hbar$. We obtain almost straight lines indicating that, as in the
 BCS case with schematic interactions, the pairing energy is proportional to  $\langle\Delta N^2 \rangle$
  justifying the use of the constraint on $\langle\Delta N^2 \rangle^{1/2}$ to generate wave functions
  with different pairing content.  As mentioned above we can see that the pairing energy interval covered by our
  calculations ranges from zero to about 50 MeV and by looking at this figure one can have an idea of the 
  behavior of  the pairing energy at each point of the plane $(q,\delta)$ in this case.

 \begin{table}[t]
\begin{center}
\begin{tabular}{c|c|c|c|c|}\cline{3-5}

\multicolumn{2}{c|}{} &HFB & HFB+PNP & PN-VAP  \\ \hline \hline

Basic approach & 1D & -470.096 &   -470.708  &  -473.066\\ \cline {2-5}

&2D  & -470.096 & -471.620   &  -473.066   \\ \cline {2-5} \hline \hline

AMP ($I=0^+$)& 1D  & -472.490 &  -472.903 &   -475.860  \\ \cline {2-5}

& 2D &  -472.491 &  -474.535& -475.860\\ \cline {2-5}  \hline \hline

AMP ($I=2^+$)& 1D  & -472.017 & -472.348  & -475.023    \\ \cline {2-5}

& 2D &  -472.017 & -473.473 & -475.025\\ \cline {2-5}  \hline \hline

SCCM (0$_{1}^+$) & 1D  & -473.522  &  -474.985 & -476.636 \\ \cline {2-5}

& 2D &  -474.137 & -475.809 & -476.865 \\ \cline {2-5} \hline \hline

SCCM (0$_{2}^+$) & 1D  & -471.282  &  -470.285 & -471.566 \\ \cline {2-5}

& 2D & -473.182  &-470.720  &  -472.232\\ \cline {2-5} \hline \hline

SCCM (2$_{1}^+$) & 1D  &  -472.541 &  -473.639 & -475.407 \\ \cline {2-5}

& 2D &  -473.096 &  -474.260& -475.598\\ \cline {2-5} \hline \hline

\end{tabular}
\end{center}
\caption{Energy minima, in MeV, in different approaches in 1D and 2D. In the first row the values obtained in the 
basic approaches (HFB, etc). In the second  row the energy obtained performing the AMP to $I=0^+$ of the corresponding
column, i.e., the number in the``HFB" column corresponds to the HFB+AMP approach, the one in the column``HFB+PNP" to the
HFB+PNAMP y the one in`PV-VAP" to the PN-VAP+PNAMP. In the third row the same as in the second one but for $I=2^+$.
 In the fourth  row the  energy of the solution of the Hill-Wheeler equation (\ref{HWeq}), (SCCM) for the $0^+_1$ state  taking as basis states the corresponding to each column as explained for the second row. In the fifth and sixth rows the same as in
 fourth row but for the $0_2^+$ and the $2_1^+$ states, respectively.}
\label{Table2}
\end{table}

As with the one dimensional case prior to the SCCM results we first  present in Fig.~\ref{2_dim_54Cr}  contour lines of the potential energy of  $^{54}$Cr as a function of the constrained parameters $(q,\delta)$  in different approaches. The bullets represent the $\delta$ values of the self-consistent  solution (HFB or PN-VAP) extracted from the one dimensional ($q$-constrained) approach and are displayed as a discussion guide.  
Since all HFB based approaches do have the same intrinsic w.f. all of them have the same bullets pattern. The same apply to all PN-VAP based approaches.  We first discuss the HFB based approaches. In the panel (1,1), here we use matrix element notation (file,column),  we display the two dimensional potential energy in the plain HFB approach, i.e., the solution of 
 Eqs.~\ref{min_E},\ref{E_prime} with $|\Phi\rangle \equiv|\phi\rangle$.  Since in this case the bullets represent the self consistent  solutions the line of the bullets is, as expected,  orthogonal to the equipotential contour lines.  As a matter of fact following the bullets we can reproduce the corresponding curve in 
 Fig.~\ref{q_plots}(a).  The gross  behavior is that for $|q|$-values larger than $~50$ fm$^2$  the equipotentials are parallel (perpendicular) to the 
 $\delta$-axis for $\delta$-values smaller (larger) than a given  $\delta_{\rm max}(q)$. That means, for a given $q$ we can modify the pairing content of the wave function up to a  given $\delta_{\rm max}(q)$ without much energy cost, i.e.,  the Hartree-Fock average potential is able to readjust itself to  compensate for the pairing energy gain, producing very soft potentials in the $\delta$ direction. However, for  $\delta$ values larger than $\delta_{\rm max}$, rather suddenly, this compensation does not take place anymore and it costs a large amount of  energy to increase the pairing correlations  of the system, 
 producing very stiff potentials. For  $|q|$-values smaller than $~50$ fm$^2$, i.e., close to the spherical shape,
there is a valley along the value $\delta\approx 2.5$, i.e., the energy increases if one tries to increase or decrease the pairing correlations.  This behavior has
probably to do with the fact that we are close to  the spherical shape and dealing with very pure configurations: four protons in the $f_{7/2}$  and two neutrons in the $p_{3/2}$ sub-shell. This constellation corresponds to a given pairing energy, and since the other sub-shells are far away in energy it takes energy to
increase or decrease the pairing correlations. 

We can get an idea about the effect of the  PNP just taking the HFB w.f.'s of the previous calculations and evaluating the PNP energy in each $(q,\delta)$ point.  This is the HFB+PNP approach which we display in panel (1,2)   of Fig.~\ref{2_dim_54Cr}.  As compared with the plain HFB results we find that the minima are somewhat deeper and slightly shifted to spherical shapes. One also observes a small shift of the equipotentials to larger $\delta$ values. In particular the absolute minimum has been shifted to a value of $\delta\approx 2$, in contrast to the plain HFB solution. We can also compare with the one dimensional plot of Fig.~\ref{q_plots}(a).  We see now in the two-dimensional plot that the kink that appears around  $q\approx 120$ fm$^2$ will not be there if we shall follow the one-dimensional path perpendicular to the equipotentials. In panel (1,3) we present
the results of the PN-VAP approach. Here we observe an enhancement of the effects mentioned with respect to the HFB+PNP plot, namely a shift of the equipotentials to higher $\delta$ values and more pronounced minima.  It is also worthwhile to observe that the 1D solution in this approach represented by the bullets   (see also Fig.~\ref{q_plots}(a))  goes perpendicular to the equipotentials in the 2D surface. 

\begin{figure*}[t]
\begin{center}
\includegraphics[angle=0,scale=0.7]{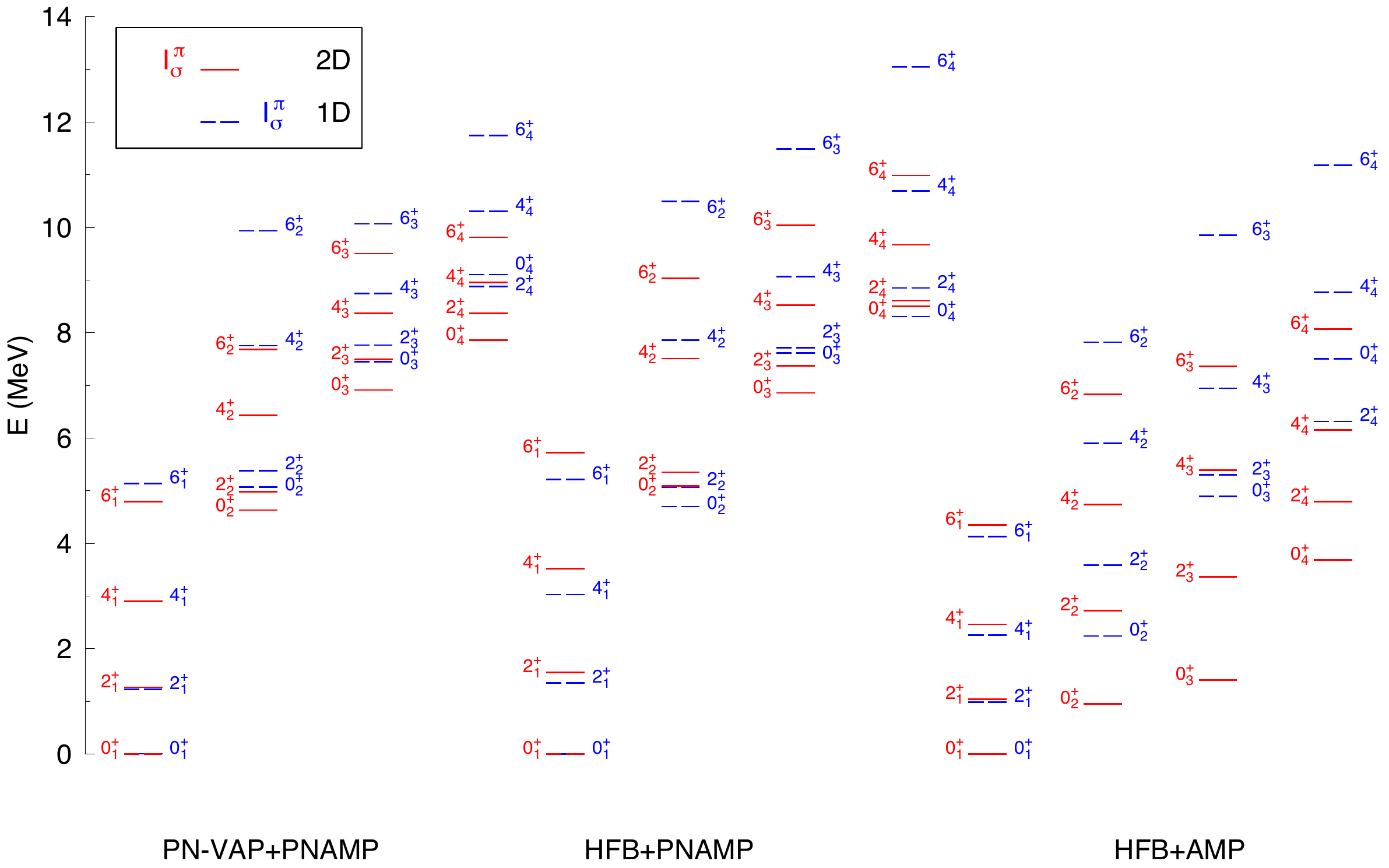}
\end{center}
\caption{ (Color online) Spectra of $^{54}$Cr in different approaches.  In the bottom of each spectrum the approach used to generate the
Hill-Wheeler basis is indicated. Dashed lines (quantum numbers at the right hand side) correspond to 1D
calculations with the quadrupole moment as generator coordinate. Continuous lines (quantum numbers at the left hand side) correspond to 2D
calculations with $q$ and $\delta$ as generator coordinates.}\label{spec_Cr}
\end{figure*}

We now discuss the effect of the AMP (for $I = 0 \hbar$) on top of the discussed approximations. That means we take now the intrinsic w.f. determined in the HFB and the PN-VAP
and calculate the AMP and/or PNP energy  in the plane $(q,\delta)$. In panel  (2,1) we display the HFB+AMP results. As compared with the HFB approach ( see panel (1,1) of Fig.~\ref{2_dim_54Cr}),  we find a softening of the equipotentials for large $q$ values and the reinforcement of both minima. Interestingly the AMP on top of the HFB does not change the position of the minima in the $(q,\delta)$ plane.  In panel (2,2) we present the HFB+PNAMP results, as compared with the HFB+PNP approach (panel (1,2)), we observe a deepening and a shift of the minima to larger $q$-values. However, we do not see a shift in the
$\delta$ direction. In panel (2,3) the  PN-VAP+PNAMP results are shown.  They differ from the PN-VAP ones in a drift of the equipotentials to larger deformations and a deepening of the minima. Concerning the $\delta$ coordinate the positions of the minima are  the same as without AMP. A comparison between  the PN-VAP+PNAMP approach and the HFB+PNAMP one reveals that the minima of the latter one are much softer towards small $\delta$ values than the  former ones.  Finally in the third row of Fig.~\ref{2_dim_54Cr} we present the results of the  PN-VAP+PNAMP approach for higher angular momenta.  As the angular momentum increases we observe  a weakening as well as a shift towards sphericity of the oblate minimum and  a drift of the equipotentials towards smaller $\delta$ values.  This effect shows clearly that with growing angular momentum 
smaller pairing energies are preferred.

\begin{figure*}[t]
\begin{center}
\includegraphics[angle=0,scale=0.7]{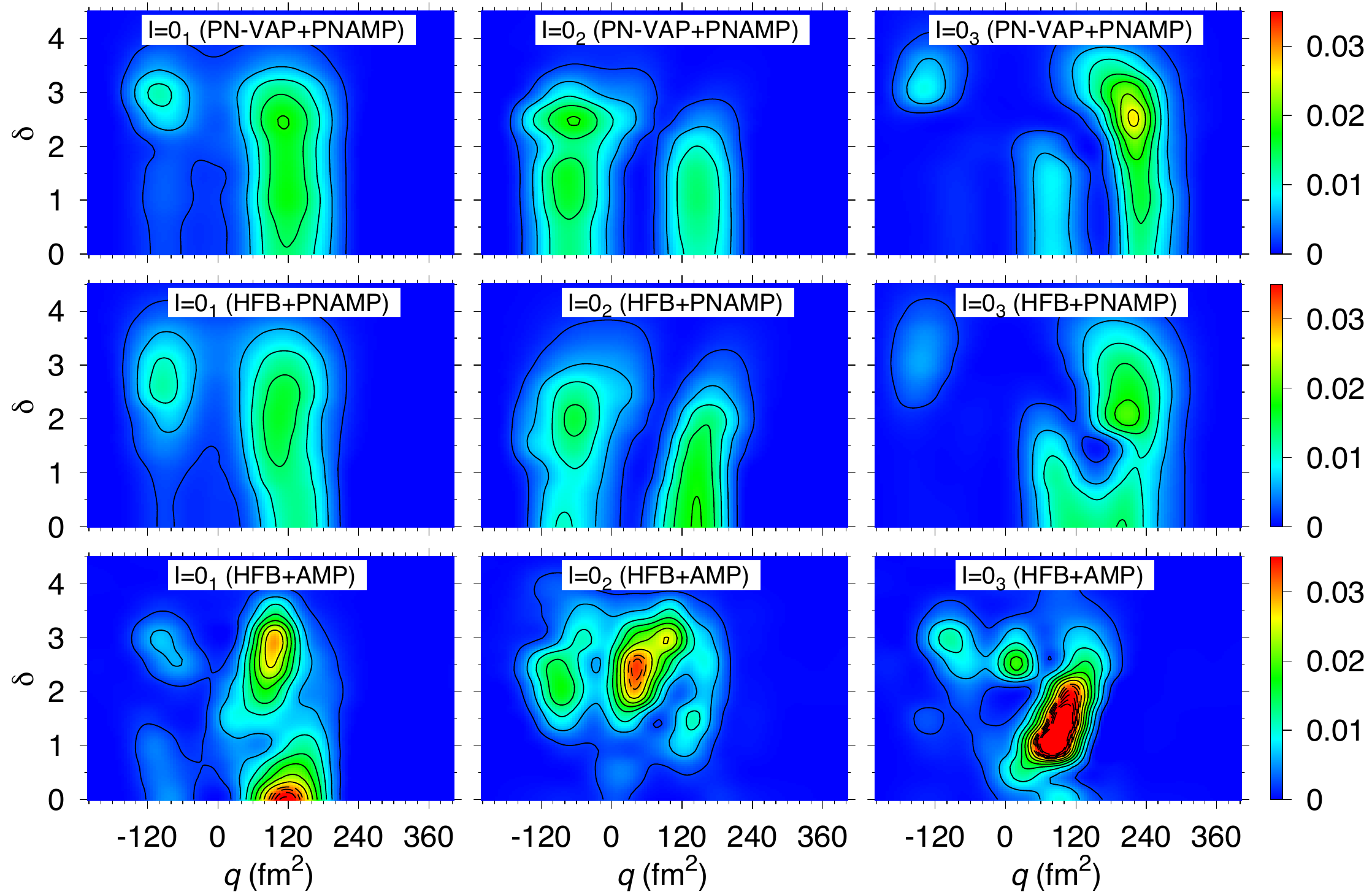}
\end{center}
\caption{(Color online)  Contour plots of the SCCM collective  wave functions in the $(\delta, q)$ plane of the three lowest $I=0 $ states in the different approaches.  In the insets of each plot the approach used to generate the
Hill-Wheeler basis is indicated.}
\label{wf_Cr}
\end{figure*}

Let us now consider the results of the SCCM calculations, for which  we have to solve the HW equation (Eq.~\ref{HWeq}). 
We again analyze  the three cases we have just discussed. To evaluate the
impact of the pairing fluctuations on the different observable we also
consider  the solutions of the HW equation in 1D with one, $(q)$,  and in 2D with two, $(q, \delta)$, generator coordinates.  We have calculated the four lowest states for each angular momentum and ordered the levels according to their quadrupole moment.  Concerning the ground states of the three approaches the middle panels  of Fig.~\ref{2_dim_54Cr} are a graphical representation of  the diagonal matrix elements of Eq.~\ref{HWeq}.  We will use these panels
to guide the analysis of the final results. However, since non-diagonal matrix elements do no follow general simple rules  it is not always easy to understand fine details of the spectra. 
The magnitude of the non-diagonal elements  and  thereby the energy gain after the solution of the HW equation depends on the approach. In general,  because of self-consistency
and the quality of the approach before the HW diagonalization we expect the smallest energy gain in the PN-VAP+PNAMP case,  a larger one in the HFB+PNAMP and the  largest one in the HFB+AMP approach.  In Table~\ref{Table2} we provide the  energy values
at the minima in different cases.  In the 2D case the energy lowering of the $0^+_1$ states in the SCCM approach (4th row) with respect to the   HFB+AMP  (2nd row) is 1.0 MeV (PN-VAP+PNAMP), 1.27 MeV (HFB+PNAMP) and 1.65 MeV (HFB+AMP). 

\begin{table*}[t]
\begin{center}
\begin{tabular}{c|c|c|c|c|c|c|}
\cline{2-7}

\multicolumn{1}{c|}{} &$0^+_{1}$ & $0^+_{2}$ & $0^+_{3}$  &$2^+_{1}$ & $2^+_{2}$ & $2^+_{3}$ \\  \cline {2-7} 
   &$\; \;E_Z\;\;\;\;\;\;E_N$ & $\; \;E_Z\;\;\;\;\;\;E_N$ & $\; \;E_Z\;\;\;\;\;\;E_N$ & $Q_{spec}$ & $Q_{spec}$& $Q_{spec}$\\ \hline \hline
PN-VAP+PNAMP-1D&  $-4.8\; \; -5.4$&$ -4.8\;\;-4.9$ &$ -4.4\;\;  -5.5$ & -29.670 & 16.845 & -49.563\\ \cline {1-7} \hline
PN-VAP+PNAMP-2D& $ -4.9\;\; -5.6$&$  -4.1\;\; -3.9$&$-4.8\;\; -6.9$ & -29.556 & 15.299 & -42.722 \\ \cline {2-7} \hline \hline
HFB+PNAMP-1D& $-2.2\; \;-2.2$& $-2.0\;\; -2.6$ & $-2.6\;\; -3.3$ &-26.997 & 14.631 & -34.864\\ \cline {1-7} \hline
HFB+PNAMP-2D& $ -3.6\; \;-5.0$&$ -2.5\;\; -2.8 $&$ -2.9\;\; -5.7$ &-27.950 & 13.824 & -40.985\\ \cline {2-7} \hline \hline
HFB+AMP-1D&$ -1.3\; \;-2.0$&$  -1.8\;\;-1.7$&$  -0.2\;\;-3.2$& -30.184 & -6.766 &-27.401   \\ \cline {1-7} \hline
HFB+AMP-2D& $ -1.7\; \; -3.0$&$ -3.3\;\; -4.1$&$ -2.9\;\;-5.5$ & -31.224 & -20.381 & -26.925  \\ \cline {2-7} \hline \hline

\end{tabular}
\end{center}
\caption{Proton and neutron pairing energies, in MeV, of the first $0^+$ states of the SCCM calculations as well as the spectroscopic quadrupole moments, in e$^2$fm$^2$, of the first $2^+$ states. In each row  the approach used to generate the
Hill-Wheeler basis is indicated.}
\label{Table3}
\end{table*}

In  Fig.~\ref{spec_Cr} we display the  spectra in the different approaches. We start with the SCCM results for the PN-VAP+PNAMP case.
When we compare the one dimensional (1D) with the two dimensional (2D) spectra we observe that the general behavior is
to compress the otherwise too much stretched 1D spectrum. The general tendency is that the lowering of the states increases with the excitation energy and with the angular momentum. The reason for this behavior, as mentioned above, is the variational principle.    To solve the HW equation in
1D  the intrinsic w.f.'s of the bullets displayed in panel (1,3) of Fig.~\ref{2_dim_54Cr} are needed, and  for the 2D the ones of the bullets plus all the others. Because of self-consistency the 1D path of the bullets goes along minima and saddle points. One should also notice that in the frame of the cranking model the condition $\langle\hat{J_x}\rangle=\sqrt{I(I+1)}$ for the case $I=0\;\hbar$ coincides with the variational principle of Eq.~\ref{min_E}. This is the reason why the bullets ``almost'' provide the right
path along the PN-VAP+PNAMP potential energy surface for $I=0\;\hbar$ in panel (2,3)  of Fig.~\ref{2_dim_54Cr}.  
One should notice, for example,  that for $I=6$, see panel (3,3) this is not the case anymore.
 However since the Yrast band is rather rotational the energy
gain of the $I=6_1^+$ state in the 2D case as compared with the 1D is not as large as for $6_2^+, 6_3^+$ or $6_4^+$ states.  Based on the comments above we expect the deviation of the 2D and 1D results  to be such that the smallest  energy gain takes place for the ground state. Furthermore,  larger deviations occur for non-Yrast states  $I_2, I_3$ etc, in such a  way that the larger the angular momentum and the energy,  the larger the deviation.
We now consider the SCCM spectra for the HFB+PNAMP case. In this case the 1D results resemble the  PN-VAP+PNAMP ones to some extend.
The 2D results however behave in a different way as compared to the PN-VAP+PNAMP.  The main discrepancy is the fact that most of the 2D states lie below the corresponding 1D ones and  in particular the ground state band is more stretched in 2D than in 1D.  The reason for this is  the lack of self-consistency clearly seen in panels (1,2) and (2,2).  Here we observe that the bullets do not proceed through the minima and one cannot take for granted that the  ground state is the one
with the lowest energy gain, see Table~\ref{Table2}. Furthermore since in the 1D calculations the pairing energy is small at the minimum the moment of inertia is large producing a more compressed spectrum than the 2D in  which the minima have stronger  pairing correlations.

Our last case is the SCCM results for the HFB+AMP approach.  The first observation is the general compression of the 1D and the 2D
calculations.  If we compare the three 1D calculations we find that  HFB+AMP are the more squeezed ones.  The reason can
be easily seen comparing the panels (2,1),  (2,2)  and  (2,3) of Fig.~\ref{2_dim_54Cr} and  Fig.~\ref{q_plots}.  There we can see that the minima in the HFB+AMP solutions, i.e., the place where the collective w.f.'s concentrate do have much less pairing correlations than in the minima of the other approaches, i.e., larger moments of inertia.  In particular,  we can  also compare in Table~\ref{Table3} the pairing energy of the $0^+_1$ states in the 1D approaches.
When we include the pairing fluctuation, since we are dealing with self-consistent calculations, though in a smaller proportion than in the case of the PN-VAP+PNAMP,  we expect a compression of the spectrum. The compression, however, is huge as compared with that case and the HFB+PNAMP one.  Since the only difference between  the HFB+AMP and the HFB+PNAMP approaches is the PNP we conclude that this ingredient is very relevant. 
The fact that the Yrast band is more compressed in the 1D than in the 2D at variance with the PN-VAP+PNAMP approach
is obviously related to the magnitude of the non-diagonal matrix elements as compared with the diagonal ones. In Table~\ref{Table3} we can observe larger pairing correlations for the ground state in the 2D than in the 1D calculations.

We can
have a clearer interpretation of the results looking at the collective wave functions.  In Fig.~\ref{wf_Cr} we display the
collective wave functions of the $0^+_1, 0^+_2$ and $0^+_3$ states in the three theoretical approaches.
The  SCCM w.f's  of the PN-VAP+PNAMP case displayed in the upper panels, are easy to understand. Looking at the panel (2,3) of Fig.~\ref{2_dim_54Cr} we find two minima and
a softening of the contour lines at $q\approx 240$ fm$^2$.  At these positions we find maxima in our wave functions.
The $0^+_1$ state has a strong peak on the prolate side which is rather soft in the $\delta$ direction and a smaller one on the oblate side. The $0^+_2$ state  has a peak on the oblate side,   with a two hump structure in the $\delta$ degree of freedom, and a weaker one on the prolate side. The  $0^+_3$ state has a big peak at a rather high $q$ value and two smaller ones.  As a matter of fact a comparison with the 1D results in the right panel of Fig.~\ref{q_plots} shows that the collective 2D
w.f.'s peaks at $q$ values close to the corresponding 1D ones.  The main characteristic of the collective PN-VAP+PNAMP 
w.f.'s is that they are very soft in the $\delta$ direction. Concerning the HFB+PNAMP w.f.'s, see second row of Fig.~\ref{wf_Cr}, they have the general appearance 
as the PN-VAP+PNAMP though they are not that soft in the $\delta $ direction. The $0^+_3$ state is  presenting 
more mixing on the prolate side than the corresponding PN-VAP+PNAMP one.  The big difference, as expected, appears in the
behavior of the  HFB+AMP collective w.f.'s. We find the strength more concentrated than in the other two cases and we find
much more mixing. In particular the prolate and oblate peaks of the $0^+_1$ and $0^+_2$ states in the PNP approaches  do mix in the HFB+AMP leading to a concentration of probability at small prolate shapes. The same applies to the $0^+_3$ state, we find an unnatural mixing of the different minima giving rise to a strange accumulation of the strength providing the collapse of the spectrum as we have found in Fig.~\ref{spec_Cr}.  These results could be an indication that the configuration mixing calculations with w.f.'s belonging to weak and strong pairing regimes do not perform properly if we treat the particle number symmetry at first order of the Kamlah approach.

 As mentioned above variational approaches are specially suited for the description of the ground state. We can study
 the convergence of the ground state energy with the improving  approaches.  As we can see in Table~\ref{Table2}, the energy in the basic approach, i.e., the plain HFB, is $-470.096$ MeV.  If we now consider self-consistently the particle number symmetry, i.e.,  the PN-VAP approach, we obtain a lowering in the energy of 2.97 MeV. The  projection of the PN-VAP wave function to zero angular momentum provides an additional decrease of 2.79 MeV. The mixing of the quadrupole degree of freedom  gives a further lowering of 0.78 MeV and,  finally, the additional consideration of the pairing fluctuations  contributes with 0.23 MeV.  These results
 indicate a good convergence of the ground state energy.

One could also rise the question about pairing vibrations. Looking to the collective w.f. of the PNP approaches we do not find clear evidence of genuine pairing vibrations.  What one observes is a strong softening of the peaks of the w.f.'s with the  pairing degree of freedom. 
In Table~\ref{Table3} we present the pairing energies of the three lowest $0^+$ states in the three approaches and
in the 1D and 2D cases.  In the PN-VAP+PNAMP case the largest values are obtained and no big difference between the 1D and 2D cases is observed.  Smaller absolute values are found in the HFB+PNAMP approach, a large increase in the neutron
pairing energies is observed for the $0^+_1$ and $0^+_3$ states. In the HFB+AMP approach one observes that
the pairing energies in the $0^+_3$ are much larger in the 2D case than in the 1D case. In a boson theory \cite{Brog_rev}
a pairing vibration is interpreted as a two boson state and one could be tempted to conclude that this state is of such
character.
This nucleus is probably not a good candidate to look for pairing vibrations and further work must be done in other regions
of the nuclide  table.

 The spectroscopic quadrupole moments of the three lowest $2^+$ states are also shown in Table~\ref{Table3}. In the
SCCM  approaches with PNP the $2^+_1$ and $2^+_3$ states are prolate and the $2^+_2$ oblate, both approaches providing
comparable results though the PN-VAP+PNAMP ones are somewhat larger. The SCCM approach with HFB+AMP w.f.'s predicts
prolate deformation for the three states and larger deviation from the PN-VAP+PNAMP one.  In general for this nucleus we do not find  big changes between the 1D and 2D quadrupole moments.

In conclusion, for the first time we have considered simultaneously pairing and quadrupole fluctuations in the
framework of the symmetry conserving configuration mixing approach with effective forces. In the studied nucleus,
$^{54}$Cr,  we find a large effect of the pairing fluctuations on the energies of the excited states. We find that particle number projection 
is very  relevant to obtain physical results. We also obtain a strong dependence on the way the intrinsic basis is
generated, the variation after projection approach providing the most consistent way. The spectra obtained without
particle number projection provide very strong mixing causing an unreasonable compression of the spectrum when
pairing fluctuations are included. In this nucleus no clear evidence of genuine pairing vibrations is found but a 
strong softening of the quadrupole modes with the pairing degrees of freedom. Neither we  find  a strong influence of the pairing fluctuations on the
quadrupole moments.

\vspace{1cm}
The authors acknowledge financial support from the Spanish Ministerio de  Ciencia e Innovaci\'on under contracts  FPA2009-13377-C02-01, by the Spanish Consolider-Ingenio 2010 Programme CPAN (CSD2007-00042). N.L.V acknowledges a scholarship of the Programa de Formaci\'on de Personal  Investigador (Ref. BES-2010-033107). T.R.R. acknowledges support from Programa de Ayudas para Estancias de Movilidad Posdoctoral 2008 and Helmholtz International Center for FAIR program.

\section{Appendix}
 As a guide to the reader we put together of all the acronyms
  in the order of appearance in the text.
\begin{description}
\item[HF]     Hartree Fock
 \item[BCS]    Bardeen Cooper Schrieffer
\item[HFB]    Hartree Fock Bogoliubov
\item[BMFT] Beyond Mean Field Theory 
\item[SCCM] Symmetry Conserving Configuration Mixing
\item[GCM] Generator Coordinate Method  
\item[PNP] Particle Number Projection 
\item[AMP]   Angular Momentum Projection
\item[PAV]   Projection After Variation  
\item[VAP]    Variation After Projection
\item[PN-VAP]  Particle Number Variation After Projection 
\end{description}

\end{document}